\documentclass[aps,prc,preprint,groupedaddress,nofootinbib]{revtex4-2}

\usepackage{graphicx}
\usepackage{amsmath,bm,ascmac,amssymb,calc}
\usepackage{CJK}

\begin{document}
\begin{CJK*}{UTF8}{}
\CJKfamily{min}


\title{Magnetic dipole moments adjacent to doubly-magic nuclei
  in self-consistent mean-field theory
  with realistic spin-isospin and tensor forces}


\author{H.~Nakada$^{1,2}$}
\email[E-mail:\,\,]{nakada@faculty.chiba-u.jp}
\author{H.~Iwata$^3$}

\affiliation{$^1$ Department of Physics, Graduate School of Science,
 Chiba University,\\
Yayoi-cho 1-33, Inage, Chiba 263-8522, Japan}
\affiliation{$^2$ Research Center for Nuclear Physics, The University of Osaka,\\
  Mihogaoka 10-1, Ibaraki, Osaka 567-0047, Japan}
\affiliation{$^3$ Department of Physics,
 Graduate School of Science and Engineering,
 Chiba University,\\
Yayoi-cho 1-33, Inage, Chiba 263-8522, Japan}


\date{\today}

\begin{abstract}
  Magnetic dipole ($M1$) moments in nuclei neighboring the doubly-magic core
  are investigated by the self-consistent mean-field (SCMF) approaches
  that allow for the breaking of the time-reversal symmetry.
  By the SCMF calculations with the M3Y-P6 interaction,
  which keeps realistic spin-isospin and tensor channels,
  the $M1$ moments are well reproduced,
  particularly those in the nuclei adjacent to $jj$-closed magicity.
  The results are in better agreement with the data
  than those with the Gogny-D1S interaction,
  slightly better than those of UNEDF1
  supplemented by a spin-isospin channel adjusted
  to the $M1$ moments themselves,
  and comparable to the shell-model results
  with the chiral effective-field-theory ($\chi$EFT) interaction.
  Analyses via quadrupole moments, occupation numbers
  and the lowest-order perturbation theory
  elucidate the cooperative effects of quadrupole deformation
  and spin correlation on the displacement from the Schmidt values,
  which has been known in terms of the quenching of the spin matrix elements.
  It is shown that a significant portion of the spin correlation is carried
  by the spin-isospin and tensor channels in the effective interaction.
  However, while agreement is remarkable at $^{131}$Sn$^m$, $^{133}$Sn
  and $^{209}$Pb,
  discrepancies remain at the $Z=\mathrm{odd}$ nuclei
  $^{133}$Sb, $^{207}$Ti and $^{209}$Bi,
  as in the $\chi$EFT-based shell-model results.
\end{abstract}


\maketitle
\end{CJK*}



\section{Introduction\label{sec:intro}}

Nucleon's spin plays specific roles in nuclear structure.
Its coupling to the orbital angular momentum yields
the spin-orbit ($\ell s$) splitting of the single-particle (s.p.) levels,
which is a source of the magic numbers
associated with the so-called $jj$-closed shell~\cite{ref:deShalit-Talmi}.
The $\ell s$ splitting significantly depends
on the proton ($Z$) and neutron ($N$) numbers,
giving rise to the appearance and disappearance of the magic numbers
in nuclei far from $\beta$-stability~\cite{ref:SP08,ref:OGSS20}.
Moreover, nuclear weak processes,
\textit{e.g.}, $\beta$-decay, double-$\beta$-decay, and neutrino scattering,
often occur through the spin degrees of freedom (d.o.f.)~\cite{
  ref:deShalit-Talmi,ref:BM1,ref:FS01,ref:BLM15}.
Because of its relevance to astrophysics and fundamental physics,
understanding and proper treatment of nucleon spin are highly desirable.

The self-consistent mean-field (SCMF) theory
provides a promising framework
for describing nuclear properties ranging from stable to unstable
and from light to heavy nuclei
in a unified manner.
In practice, it has the ability to reproduce and predict
nuclear magic numbers all over the nuclear chart
from scratch~\cite{ref:VB72,ref:NS14}.
The SCMF theory connects the properties of finite nuclei
to those of infinite nuclear matter,
making the best use of experiments on Earth to predict nuclear properties
in astrophysical environments, such as neutron stars.
This advantage can hold for the spin properties,
as far as they are well examined in finite nuclei.
Furthermore,
while the SCMF theory is suitable for describing the lowest-energy states,
such as ground state (g.s.), intrinsic state of ground rotational band,
and isomeric state,
it is straightforwardly extended to excitations
via the random-phase approximation (RPA)~\cite{ref:PB52,ref:RS80,ref:GS81}
and nuclear scattering via the folding model~\cite{ref:DG72,ref:NI24}.
The SCMF theory also supplies a basis of extended approaches;
the generator-coordinate method~\cite{ref:RG87,ref:BHR03},
the particle-vibration-coupling model~\cite{ref:HS76,ref:CSB10},
the subtracted second RPA~\cite{ref:GGE15},
the time-dependent density-matrix theory~\cite{ref:WC85,ref:Toh20},
the hybrid configuration mixing model~\cite{ref:CBB17},
and so forth.
There have been arguments for reinterpreting the nuclear SCMF theory
in terms of density-functional theory
(DFT)~\cite{ref:PS91,ref:Ring16,ref:Colo20}.
The SCMF equations have the identical form to the Kohn-Sham (KS)~\cite{ref:KS65}
or Kohn-Sham--Bogoliubov-de Gennes equation~\cite{ref:OGK88}.
The so-called \textit{ab initio} approaches have progressed,
in which many-body correlations are taken into account
using the coupled-cluster method~\cite{ref:HPHD14},
the self-consistent Green's function method~\cite{ref:DB04,ref:Soma},
the similarity-renormalization-group (SRG) method~\cite{ref:HBMS16},
and others.
The KS theory may conceptually be an alternative approach to them,
which takes account of many-body correlations in the energy functional (EF).
The KS theory may be extended
so that principal variables could be constituted
by a class of physical quantities
represented by the one-body density matrix,
whether it contains $\rho(\mathbf{r})$ or not~\cite{ref:Nak23}.
It seems sensible to identify the nuclear SCMF theory,
which could cover the spin properties,
with the (generalized) KS theory for nuclei,
though we keep referring to it as SCMF theory in this paper to avoid confusion.

The SCMF theory needs an effective interaction as its input,
which encompasses the effects of many-body correlations
and corresponds to the EF of the KS theory.
It is never a trivial task to establish the effective interaction or EF.
A popular method is fitting parameters to measured quantities
with respect to nuclear structure~\cite{ref:UNEDF1,ref:UNEDF2},
after assuming the form of EF, typically the Skyrme form.
However, since the spin d.o.f. tend to be masked
as an effect of the Pauli principle and nucleonic interaction,
it is not easy to fix the spin-dependent channels
via fitting~\cite{ref:UNEDF1,ref:UNEDF2}.
As the spin operator is not invariant
under the time-reversal transformation $\mathcal{T}$,
this problem is related to the $\mathcal{T}$-symmetry
in the mean field~\cite{ref:HR88,ref:FP89,ref:YCM06,ref:SDBG22}.
The $\mathcal{T}$-symmetry usually holds
in the g.s.'s of even-even nuclei.
The spin properties are more visible in odd nuclei or proper excitations.
It should be kept in mind that the non-central channels in the interaction
may significantly influence the spin properties of finite nuclei.

The magnetic dipole ($M1$) moment is an observable sensitive to spin.
In particular, those of nuclei neighboring the doubly-magic core
supply a good testing ground for the theoretical description
of the spin properties.
For the single-particle or single-hole states,
the $M1$ moments are governed by the orbital of the last nucleon~\cite{
  ref:deShalit-Talmi,ref:BM1,ref:Blatt-Weisskopf},
reaching the Schmidt values [see Eq.~\eqref{eq:mm-Sch}].
However, the measured $M1$ moments sizably deviate from the Schmidt values
in most nuclei~\cite{ref:Blatt-Weisskopf},
implicating quenching of the spin matrix elements.
By microscopically investigating various effects in the perturbation theory,
the main source of quenching has been attributed
to core polarization (CP)~\cite{ref:Tow87,ref:ASBH88,ref:LM18},
after a history of disputes.
Thus, the degree of quenching reflects the stiffness of the doubly-magic core,
and is influenced by spin-dependent channels of the interaction.
As the $M1$ moment is basically represented by the one-body density matrix,
it can be taken as a principal variable of the KS approach.
In this context,
it has been shown that deformation and $\mathcal{T}$-symmetry breaking
cooperatively induce moderate quenching~\cite{ref:HR88,ref:FP89,ref:YCM06,
  ref:SDBG22}.
The approach has been extended to open-shell nuclei~\cite{ref:BE17,ref:PHGM21,
  ref:VGMB22,ref:BDDK23}.

The $M1$ moments in the vicinity of the doubly-magic nuclei
have also been investigated
by applying the in-medium SRG treatment
of the nucleonic interaction and operator
based on the chiral effective field theory ($\chi$EFT)~\cite{ref:MCSB24}.
Roles of the CP in the quenching have been manifested
from a microscopic standpoint,
and corrections from the meson-exchange current (MEC) have been argued.

As well as the moments,
$M1$ and Gamow-Teller (GT) excitations are subject to the quenching
of the spin matrix elements~\cite{ref:Tow87,ref:ASBH88}.
Under an appropriate condition,
nuclear reactions pick up the spin d.o.f.,
possessing rates proportional to the $M1$ or GT excitations.
The dominant role of the CP in the quenching
has been supported by the nuclear reaction experiments~\cite{ref:WSOO97}.
It has also been suggested that the quenching occurs in the isovector component,
leaving the isoscalar component almost unquenched~\cite{ref:MTNA15}.

In nucleonic interactions,
spin-dependence partly originates from the central channel
in the one-pion exchange potential (OPEP).
The OPEP channel is not explicitly included in most SCMF calculations,
although some of its effects might be substituted in other forms.
It deserves investigating the spin-dependent channels of effective interaction
in connection with the OPEP.
There is no doubt that the tensor force is contained
in the nucleonic interaction,
whose substantial part also originates from the pion exchange.
While the tensor force is not explicitly included
in the usual SCMF calculations,
there have been some SCMF approaches
in which the tensor force is incorporated~\cite{ref:SBF77,ref:CL98,
  ref:OSFG05,ref:BDOA06,ref:CSFB07,ref:Sky-TNS,ref:SC14,ref:Nak20}.
The tensor force affects the interaction between nucleons on specific orbits,
and this effect gives rise to
$Z$- and $N$-dependence of the shell structure~\cite{ref:OGSS20,ref:OSFG05,
  ref:BDOA06,ref:Nak10,ref:NSM13,ref:NS14,ref:SC14,ref:NI24b}.
Indeed, the prediction of magic numbers is greatly improved
by incorporating a realistic tensor force
into the effective interaction~\cite{ref:NS14}.
As the tensor force influences spin properties of nuclei,
it is of interest to investigate the effects of the realistic tensor force
on the quenching in the $M1$ moments.
Because the contribution of the tensor force
vanishes in homogeneous nuclear matter,
it is vital to separate the tensor-force effects
on the properties of finite nuclei
from those of the central channels
when extrapolating spin properties on Earth to astrophysical environments.

One of the authors (H.N.) has developed effective interactions
for SCMF approaches~\cite{ref:Nak20,ref:Nak03}
by modifying the Michigan-three-range-Yukawa (M3Y)
interaction~\cite{ref:M3Y,ref:M3Y-P}.
The M3Y interaction was originally obtained from the $G$-matrix.
In the M3Y-P$n$ interactions,
where $n$ is an integer specifying the parameter-set,
density-dependent terms are added to reproduce the saturation,
and the values of several parameters are phenomenologically modified.
In the M3Y-P6 interaction~\cite{ref:Nak13},
the central channel of OPEP
and the tensor force derived from the $G$-matrix are kept unchanged
with no phenomenological modification.
With the inclusion of these spin-isospin and tensor forces,
the M3Y-P6 interaction provides a good description of the nuclear magic numbers
up to the $Z$- and $N$-dependence~\cite{ref:NS14,ref:Nak20},
and the s.p. potential up to as high as
$\approx 80\,\mathrm{MeV}$~\cite{ref:NI24}.
The tensor force in M3Y-P6 is scrutinized in the $N$-dependence
of the s.p. level spacing between $p0d_{3/2}$ and $p1s_{1/2}$
from $^{40}$Ca to $^{48}$Ca~\cite{ref:NSM13,ref:Nak20}.
The significance of the tensor force in the $M1$ excitations
has been shown for the $M1$ excitations at $^{208}$Pb~\cite{ref:SHOT08}.
The connection to bare nucleonic interaction
and the reasonable agreement with available data
supply a good reason to regard this effective interaction
as a \textit{semi-realistic} one.
In particular, the spin-isospin and tensor channels are considered realistic.
In this work, we apply the M3Y-P6 interaction
to the $M1$ moments adjacent to the doubly-magic nuclei.

For comparison, we also employ the Gogny-D1S interaction~\cite{ref:D1S}.
Successfully describing many nuclear properties,
it is one of the standard phenomenological SCMF interactions
and is useful to inspect the roles of the realistic spin-isospin
and tensor forces.
Comparison is also made with the results given in Refs.~\cite{ref:SDBG22}
and \cite{ref:MCSB24}.

\section{Theoretical backgrounds and numerical setups\label{sec:cal}}

\subsection{Symmetries in mean-field calculations\label{subsec:sym}}

The total angular momentum $J$ is a half-integer in odd-$A$ nuclei.
Therefore, it is impossible to keep the $\mathcal{T}$-symmetry
with a single pure state $|\Phi\rangle$,
because $\mathcal{T}^2|\Phi\rangle=-|\Phi\rangle$
and thereby $\mathcal{T}|\Phi\rangle$ is necessarily orthogonal
to $|\Phi\rangle$.
The $\mathcal{T}$-symmetry should be broken through the spin d.o.f.,
supplying a good opportunity to investigate spin properties.
However, computations without imposing $\mathcal{T}$-symmetry
are often demanding,
particularly when the semi-realistic interactions are applied.
On the other hand,
it is expected that the spherical symmetry is not severely broken
in the nuclei neighboring the doubly-magic core.
To reduce computational cost,
it is reasonable to take into account symmetry breaking to a minimal extent.
We here assume the axial symmetry
under which the $z$-component of the total angular momentum ($J_z$)
is conserved,
the parity ($\mathcal{P}$) symmetry,
and the symmetry with respect to the product of $\mathcal{R}$ and $\mathcal{T}$,
where $\mathcal{R}\,(:=e^{-i\pi J_y})$ denotes the reflection,
as discussed in Appendix~\ref{app:T-sym}.
Note that, because $(\mathcal{RT})^2|\Phi\rangle=|\Phi\rangle$~\cite{ref:BM1},
the $\mathcal{RT}$-symmetry or $\mathcal{RPT}$-symmetry can be imposed
even when the $\mathcal{T}$-symmetry is violated.
Under the phase convention adopted here,
we have $(\mathcal{RPT})|\Phi\rangle=|\Phi\rangle$.

\subsection{Setups for mean-field calculations\label{subsec:setup}}

We carry out Hartree-Fock (HF) calculations
using the M3Y-P6 interaction,
employing the Gaussian expansion method (GEM)~\cite{ref:NS02,ref:Nak06,
  ref:Nak08}.
The basis functions are given in Ref.~\cite{ref:Nak08}.
As deformation gives rise to mixing of the orbital angular momentum $\ell$
in the s.p. states,
the space is truncated by its maximum value $\ell_\mathrm{cut}$.

The computer code has been newly extended
for mean-field calculations assuming the $J_z$, $\mathcal{P}$
and $\mathcal{RT}$ symmetries,
by which the spin d.o.f. can be active.
Although the present study is limited to the HF results,
the code has been adapted to the Hartree-Fock-Bogoliubov calculations,
as discussed in Appendix~\ref{app:axial}.
The convergence for $\ell_\mathrm{cut}$ against deformation
has been inspected in Ref.~\cite{ref:Nak08}.
The code adopts $\ell_\mathrm{cut}=7$ as a default value.
The nuclei under investigation do not gain strong deformation,
as will be confirmed from the $Q_p$ values
shown in Sec.~\ref{subsec:results_E2mom},
and $\ell_\mathrm{cut}=7$ is sufficient for the $N\leq 51$ nuclei.
For the $N\geq 81$ nuclei, we adopt $\ell_\mathrm{cut}=8$
to take care of an s.p. state dominated by the $0i$ component,
in which the $\ell=8$ component may be admixed to some extent.

For the odd-$A$ nuclei under investigation,
several states having different $J_z$ values lie with close energies,
which correspond to magnetic substates.
We compute all these states and adopt the state with the lowest energy
among them.

\subsection{Magnetic dipole moments\label{subsec:def_M1mom}}

The nuclear $M1$ moment operator is,
\begin{equation}
  \mu^{(1)}_0 = \sum_{\tau=p,n} \sum_{i\in\tau} \big(g_{\ell,\tau}\,\ell_{i,z}
  + g_{s,\tau}\,s_{i,z}\big)\,,
\label{eq:mm-op}\end{equation}
where we denote orbital and spin angular momenta of the $i$-th nucleon
by $\boldsymbol{\ell}_i$ and $\mathbf{s}_i$.
The s.p. $g$-factors on Eq.~\eqref{eq:mm-op} are
$g_{\ell,p}=1$, $g_{\ell,n}=0$,
and $g_{s,\tau}=2\mu_\tau$ with the measured $M1$ moment of a single nucleon
$\mu_\tau$~\cite{ref:PDG20}.
In the energy eigenstate of the $^A Z$ nuclide
$\big|\Psi_{JM}(^A Z)\big\rangle$,
the total angular momentum $J$ is a good quantum number.
At the g.s.'s of odd-$A$ nuclei,
$J\ne 0$ and its magnetic substates labeled by $M$ are degenerate.
The expectation value of $\mu^{(1)}_0$ at $\big|\Psi_{JM}(^A Z)\big\rangle$
linearly depends on $M$,
owing to the Wigner-Eckart theorem,
\begin{equation}\begin{split}
  \big\langle\Psi_{JM}(^A Z)\big|\mu^{(1)}_0\big|\Psi_{JM}(^A Z)\big\rangle
  &= \frac{\big\langle\Psi_J(^A Z)\big|\big|\mu^{(1)}
    \big|\big|\Psi_J(^A Z)\big\rangle}{\sqrt{2J+1}}\,
  \big(J\,M\,1\,0\big|J\,M\big) \\
  &= M\,\frac{\big\langle\Psi_J(^A Z)\big|\big|\mu^{(1)}
    \big|\big|\Psi_J(^A Z)\big\rangle}
  {\sqrt{J(J+1)(2J+1)}}\,.
\end{split}\label{eq:mm-M}\end{equation}
The $M1$ moment is defined for the $M=J$ state,
\begin{equation}\begin{split}
  \mu(^A Z)
  :&= \big\langle\Psi_{JJ}(^A Z)\big|\mu^{(1)}_0\big|\Psi_{JJ}(^A Z)\big\rangle \\
  &= \sqrt{\frac{J}{(J+1)(2J+1)}}\,
  \big\langle\Psi_J(^A Z)\big|\big|\mu^{(1)}\big|\big|\Psi_J(^A Z)\big\rangle\,.
\end{split}\label{eq:mm-J}\end{equation}
Instead of computing the reduced matrix element,
Eq.~\eqref{eq:mm-J} can be rewritten as
\begin{equation}
  \mu(^A Z) = J\,\frac{\big\langle\Psi_{JM}(^A Z)\big|\mu^{(1)}_0
    \big|\Psi_{JM}(^A Z)\big\rangle}
  {\big\langle\Psi_{JM}(^A Z)\big|J_z\big|\Psi_{JM}(^A Z)\big\rangle}\,,
\label{eq:mm-ratio}\end{equation}
with the $J_z$ operator on the denominator of the rhs given by
\begin{equation}
  J_z = \sum_i \big(\ell_{i,z} + s_{i,z}\big)\,.
\label{eq:Jz-op}\end{equation}

In the SCMF solution $|\Phi(^A Z)\rangle$,
in which the full rotational symmetry is spontaneously broken,
degeneracy with respect to $M$ is lost
and we choose the state having the lowest energy among them.
Within the context of the KS theory,
$\big\langle\Phi(^A Z)\big|\mu^{(1)}_0\big|\Phi(^A Z)\big\rangle$
corresponds to the $M1$ moment of the g.s.
if the $M1$ moment belongs to principal variables,
even though the wave function $\big|\Phi(^A Z)\big\rangle$
is not completely physical.
However, to compare it with the experimental data of Eq.~\eqref{eq:mm-J},
a prescription is necessary in order to match the $M$-dependence.
Applying Eq.~\eqref{eq:mm-ratio} to the SCMF solution,
we calculate the $M1$ moment within the SCMF,
\begin{equation}
  \mu^\mathrm{MF}(^A Z) = J\,\frac{\big\langle\Phi(^A Z)\big|\mu^{(1)}_0
    \big|\Phi(^A Z)\big\rangle}
  {\big\langle\Phi(^A Z)\big|J_z\big|\Phi(^A Z)\big\rangle}\,.
\label{eq:mm-MF}\end{equation}
For the $J$ value on the rhs of Eq.~\eqref{eq:mm-MF},
we insert the $J$ value of the observed g.s. (or of the metastable state).

In contrast to the KS theory,
the wave function $\big|\Phi(^A Z)\big\rangle$ is respected
in the conventional interpretation of the SCMF theory.
Then, as $J$ should be a good quantum number in the energy eigenstates,
the $J$-projected state is considered to correspond to the eigenstate,
$\big|\Psi_{JM}(^A Z)\big\rangle \propto P_{JM}\big|\Phi(^A Z)\big\rangle$,
where $P_{JM}$ is the angular-momentum-projection operator.
The $M1$ moment calculated with the projected state is
\begin{equation}
  \mu^\mathrm{proj.}(^A Z) = J\,
  \frac{\big\langle\Phi(^A Z)\big|P_{JM}\,\mu^{(1)}_0
    P_{JM}\,\big|\Phi(^A Z)\big\rangle}
  {\big\langle\Phi(^A Z)\big|P_{JM}\,J_z\,P_{JM}\big|\Phi(^A Z)\big\rangle}\,.
\label{eq:mm-proj}\end{equation}
However, $J$-projection requires a demanding and careful computation,
particularly for odd-$A$ nuclei.
Whereas a code for the g.s.'s of even-even nuclei
has been developed~\cite{ref:AN22},
no $J$-projection code is available for the SCMF solutions in odd-$A$ nuclei
associated with the GEM at this moment.
It is a delicate question that should depend on the input,
\textit{i.e.}, effective interaction or EF,
which of the pictures relying on the wave function
or the picture provided by the KS theory is suitable.
In the vicinity of the doubly-magic core,
the breaking  of the $J$ quantum number is not quite serious.
As will be shown in Sec.~\ref{subsec:results_M1mom},
we have $\mu^\mathrm{MF}(^A Z)\approx \mu^\mathrm{proj.}(^A Z)$.
It allows us to postpone, for the time being,
answering which picture is appropriate for the present case.

For a state having one-particle or one-hole on top of the doubly-magic core
without any residual correlations,
the orbit of the particle or hole determines the $M1$ moment
because the core has no contribution.
This s.p. value is known as the Schmidt value~\cite{ref:BM1},
\begin{equation}
  \mu^\mathrm{s.p.}(\tau,\ell j) = j\bigg[g_{\ell,\tau} \pm
    \frac{g_{s,\tau}-g_{\ell,\tau}}{2\ell+1}\bigg]\quad
  \Big(\,j=\ell\pm\frac{1}{2}\,\Big)\,.
\label{eq:mm-Sch}\end{equation}
Even in the nuclei adjacent to the doubly-magic core,
the measured $M1$ moments deviate from $\mu^\mathrm{s.p.}(\tau,\ell j)$
in practice.
We shall mainly discuss displacement of the measured ($\mu^\mathrm{exp.}$)
and calculated $M1$ moments from $\mu^\mathrm{s.p.}(\tau,\ell j)$,
defining
\begin{equation}
  \varDelta\mu(^A Z) := \mu(^A Z) - \mu^\mathrm{s.p.}(\tau,\ell j)\,.
\label{eq:delta-mm}\end{equation}

To be precise,
the center-of-mass (c.m.) motion is desirable to be removed
in the orbital angular momentum $\boldsymbol{\ell}_i$
in Eqs.~\eqref{eq:mm-op} and \eqref{eq:Jz-op}
with replacing $\boldsymbol{\ell}_i$ by $\boldsymbol{\ell}'_i$,
where
\begin{equation}
  \boldsymbol{\ell}'_i = \mathbf{r}'_i \times \mathbf{p}'_i\,;\quad
  \mathbf{r}'_i=\mathbf{r}_i-\mathbf{R}\,,~
  \mathbf{p}'_i=\mathbf{p}_i-\frac{\mathbf{P}}{A}\,,~
  \mathbf{R}:=\frac{1}{A}\sum_i \mathbf{r}_i\,,~
  \mathbf{P}:=\sum_i \mathbf{p}_i\,.
\label{eq:ell-cmcorr}\end{equation}
We have confirmed that this c.m. correction is negligibly small.
In relation to the deviation from $\mu^\mathrm{s.p.}$,
the influences of the MEC
and the isobar (\textit{e.g.}, $\Delta$-$h$) excitation
have been argued.
We neglect them, yielding a brief discussion on this point
in Sec.~\ref{sec:summary}.

\subsection{Electric quadrupole moments\label{subsec:def_E2mom}}

While the main subject of this paper is the $M1$ moment,
the quadrupole moments also carry useful information
about the s.p. nature of the nuclei neighboring the doubly-magic core.
The quadrupole operator is given as
\begin{equation}
  Q^{(2)}_{0,\tau} = \sqrt{\frac{16\pi}{5}}
  \sum_{i\in\tau} r_i^{\prime 2}\,Y^{(2)}_0(\hat{\mathbf{r}'}_i)\,.
\label{eq:Q-op}\end{equation}
Here and in the following,
$r=|\mathbf{r}|$ and $\hat{\mathbf{r}}=\mathbf{r}/r$.
The electric quadrupole ($E2$) moment is
\begin{equation}\begin{split}
  eQ_p(^A Z) :&= \big\langle\Psi_{JJ}(^A Z)\big|eQ^{(2)}_{0,p}
  \big|\Psi_{JJ}(^A Z)\big\rangle \\
  &= \frac{\big\langle\Psi_J(^A Z)\big|\big|eQ^{(2)}_p
    \big|\big|\Psi_J(^A Z)\big\rangle}{\sqrt{2J+1}}\,
  \big(J\,J\,2\,0\big|J\,J\big)\,.
\end{split}\label{eq:eQp-J}\end{equation}

From the SCMF solution $|\Phi(^A Z)\rangle$,
the $E2$ moments are calculated by
\begin{equation}
  eQ_p^\mathrm{MF}(^A Z) = c_M \big\langle\Phi(^A Z)\big|eQ^{(2)}_{0,p}
  \big|\Phi(^A Z)\big\rangle\,.
\label{eq:Q-MF}\end{equation}
Since $|\Phi(^A Z)\rangle$ is selected as the lowest-energy state
among several candidates having various $J_z$ values,
we have the factor $c_M$ on the rhs of Eq.~\eqref{eq:Q-MF},
for which we adopt
\begin{equation}
  c_M = \frac{3J^2-J(J+1)}{3M^2-J(J+1)}\,;\quad
  M := \big\langle\Phi(^A Z)\big|J_z\big|\Phi(^A Z)\big\rangle\,,
\end{equation}
as indicated by the Wigner-Eckart theorem.
For $J$ in $c_M$,
the value of the observed g.s. (or of the metastable state) is inserted again.

Even when the doubly magic core is entirely spherical without polarization,
the last nucleon or hole produces a quadrupole moment.
We take this s.p. value as a reference,
since the deviation from it suggests how much the core is polarized.
The s.p. value of $Q(^A Z)$ is expressed as
\begin{equation}
  Q^\mathrm{s.p.}(n\ell j) = \pm \sqrt{\frac{16\pi}{5}}
  \langle n\ell j|r^2|n\ell j\rangle_r\,
  \frac{\langle\ell j||Y^{(2)}_0(\hat{\mathbf{r}})||\ell j\rangle}
       {\sqrt{2j+1}}\,
  \big(j\,j\,2\,0\,|\,j\,j\big)\,.
\label{eq:Q-sp}\end{equation}
To analytically estimate the radial matrix element
$\langle n\ell j|r^2|n\ell j\rangle_r$ in Eq.~\eqref{eq:Q-sp},
we employ the isotropic harmonic-oscillator wave-function
with $\hbar\omega=41.2A^{-1/3}\,\mathrm{MeV}$.
The negative sign on the rhs of Eq.~\eqref{eq:Q-sp}
applies to single-hole states.

\section{Results\label{sec:results}}

\subsection{Spin properties of effective interactions in infinite nuclear matter
  \label{subsec:prop_effint}}

Before discussing the results of the $M1$ moments,
we compare spin properties in the nuclear matter
provided by the currently used effective interactions or EFs.

We emphasize again that the M3Y-P6 interaction keeps the spin-isospin channel
of the OPEP,
which we denote by $V^{(\mathrm{C})}_\mathrm{OPEP}$,
and that it has the tensor force $V^{(\mathrm{TN})}$ derived from the $G$-matrix.
In contrast,
neither of $V^{(\mathrm{C})}_\mathrm{OPEP}$ nor $V^{(\mathrm{TN})}$
is explicitly included in most phenomenological interactions,
\textit{e.g.}, D1S,
although a part of their effects might be incorporated in an effective manner.

The spin properties predicted by each effective interaction
are typically expressed in terms of the Landau-Migdal parameters.
We compare the $g_\ell$ and $g'_\ell$ values for $\ell=0,1$
obtained from M3Y-P6 and D1S~\cite{ref:Nak03} in Table~\ref{tab:LM-param}.
Whereas experimental information is limited,
the data on the GT transition suggest
$g'_0\approx 1$~\cite{ref:SS99,ref:WIS05}\,\footnote{
In Refs.~\cite{ref:SS99,ref:WIS05},
the $g'$ values are measured in the unit $(f_\pi/m_\pi)^2$,
which should be scaled by a factor of $2.6M^\ast_0/M\sim 1.8$
when comparing with the normal values,
where $M^\ast_0$ is the $k$-mass at the Fermi momentum.
}.
It seems that D1S gives too small $g'_0$.
In the M3Y case, $g'_0$ seems consistent with the data,
to which $V^{(\mathrm{C})}_\mathrm{OPEP}$ yields
an important contribution~\cite{ref:Nak03}.

\begin{table}
\begin{center}
  \caption{Comparison of Landau-Migdal parameters $g_{0,1}$ and $g'_{0,1}$
    at the saturation point
    among effective interactions and EFs.
    The values for UNEDF1 are those adopted in Ref.~\protect\cite{ref:SDBG22}.
\label{tab:LM-param}}
\begin{tabular}{cr@{.}lr@{.}lr@{.}l}
\hline\hline
\hspace*{1cm} & \multicolumn{2}{c}{M3Y-P6~~} & \multicolumn{2}{c}{~~~D1S~~~} &
 \multicolumn{2}{c}{UNEDF1} \\ \hline
$g_0$ & ~~$0$&$272$ & ~~$0$&$466$ & ~~~~$0$&$4$ \\
$g_1$ & ~~$0$&$231$ & ~$-0$&$184$ & ~~~~$0$&$0$ \\
 \hline
$g'_0$ & ~~$0$&$970$ & ~~$0$&$631$ & ~~~~$1$&$7$ \\
$g'_1$ & ~~$0$&$157$ & ~~$0$&$610$ & ~~~~$0$&$0$ \\
\hline\hline
\end{tabular}
\end{center}
\end{table}

UNEDF1 does not originally contain spin-dependent channels.
In Ref.~\cite{ref:SDBG22},
simple terms corresponding to the delta interaction have been added
to the spin-dependent channels,
whose strengths are determined
in terms of the Landau-Migdal parameter $g_0$ and $g'_0$.
For the isoscalar term, $g_0=0.4$ is imposed.
The $M1$ moments significantly depend on the spin-isospin interaction.
Utilizing this dependence,
$g'_0$ is adjusted to the $M1$ moments, and the value $1.7$
is adopted\,\footnote{
This $g'_0$ value is not incompatible with the data
because UNEDF1 gives a large $M^\ast_0\,(\approx M)$.
}.
These $g_0$ and $g'_0$ values used together with UNEDF1
in Ref.~\cite{ref:SDBG22}
are also shown in Table~\ref{tab:LM-param}.

\subsection{Magnetic dipole moments\label{subsec:results_M1mom}}

\begin{figure}
  \includegraphics[scale=0.7]{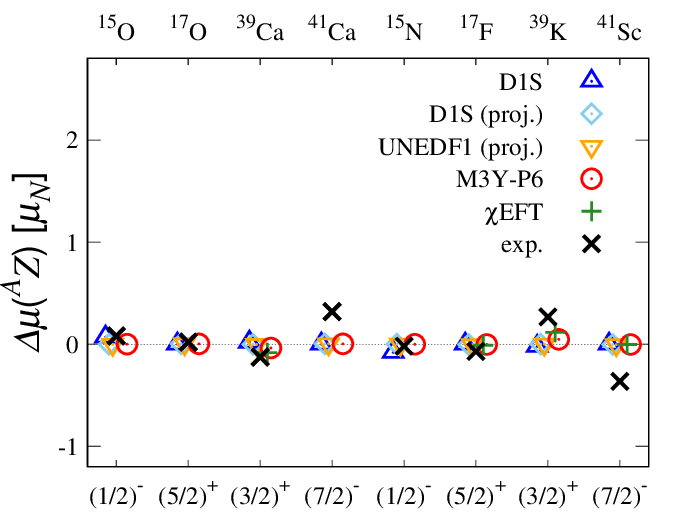}
  \caption{$\varDelta\mu$ values in nuclei adjacent to $\ell s$-closed shell.
    Nuclides and spin-parities are shown at the top and bottom of the figure.
    Red circles and blue triangles represent the present results
    with M3Y-P6 and D1S, respectively.
    Skyblue diamonds and orange inverse triangles
    are the $J$-projected results with D1S and UNEDF1
    quoted from Ref.~\protect\cite{ref:SDBG22}.
    Green pluses are $\chi$EFT results of Ref.~\protect\cite{ref:MCSB24}.
    Experimental data are taken from Ref.~\protect\cite{ref:INDC-0794}
    and shown by black crosses.
\label{fig:mu_ls-ls_A-odd}}
\end{figure}

We first depict the $\varDelta\mu$ values in nuclei
adjacent to $^{16}$O and $^{40}$Ca, in Fig.~\ref{fig:mu_ls-ls_A-odd}.
In these nuclei,
the $M1$ moments hardly deviate from $\mu^\mathrm{s.p.}$
within the lowest-order perturbation theory (see Sec.~\ref{subsec:perturb}),
because of the $\ell s$-closure at $^{16}$O and $^{40}$Ca.
Therefore, it is not surprising
that all the SCMF results yield negligibly small $\varDelta\mu$.
Although $\varDelta\mu$ is visible in the experimental data near $^{40}$Ca,
it still stays small.

\begin{figure}
  \includegraphics[scale=0.7]{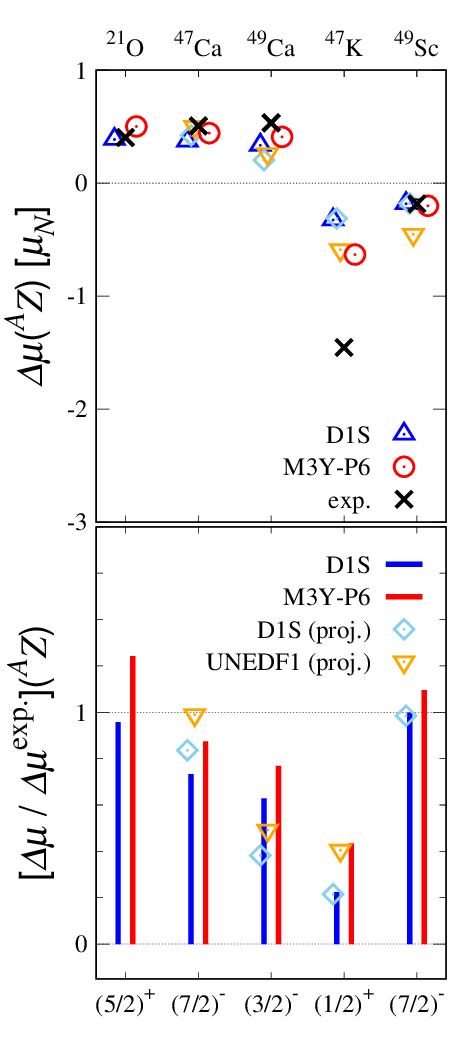}
  \caption{Upper panel:
    $\varDelta\mu$ values in nuclei adjacent to the doubly-magic nuclei,
    having $\ell s$-closed $Z$ and $jj$-closed $N$, and \textit{vice versa}.
    See Fig.~\protect\ref{fig:mu_ls-ls_A-odd} for conventions.
    Experimental data are taken from Refs.~\protect\cite{ref:INDC-0794}
    and \protect\cite{ref:IGIT23} (for $^{21}$O).
    Lower panel:
    Calculated $\varDelta\mu$ values relative to $\varDelta\mu^\mathrm{exp.}$.
    Red and blue bars are the current results with M3Y-P6 and D1S,
    respectively.
    The $J$-projected results with D1S and UNEDF1
    quoted from Ref.~\protect\cite{ref:SDBG22} are also presented
    by skyblue diamonds and orange inverse triangles, for reference.
\label{fig:mu_ls-jj_A-odd}}
\end{figure}

\begin{figure}
  \includegraphics[scale=0.7]{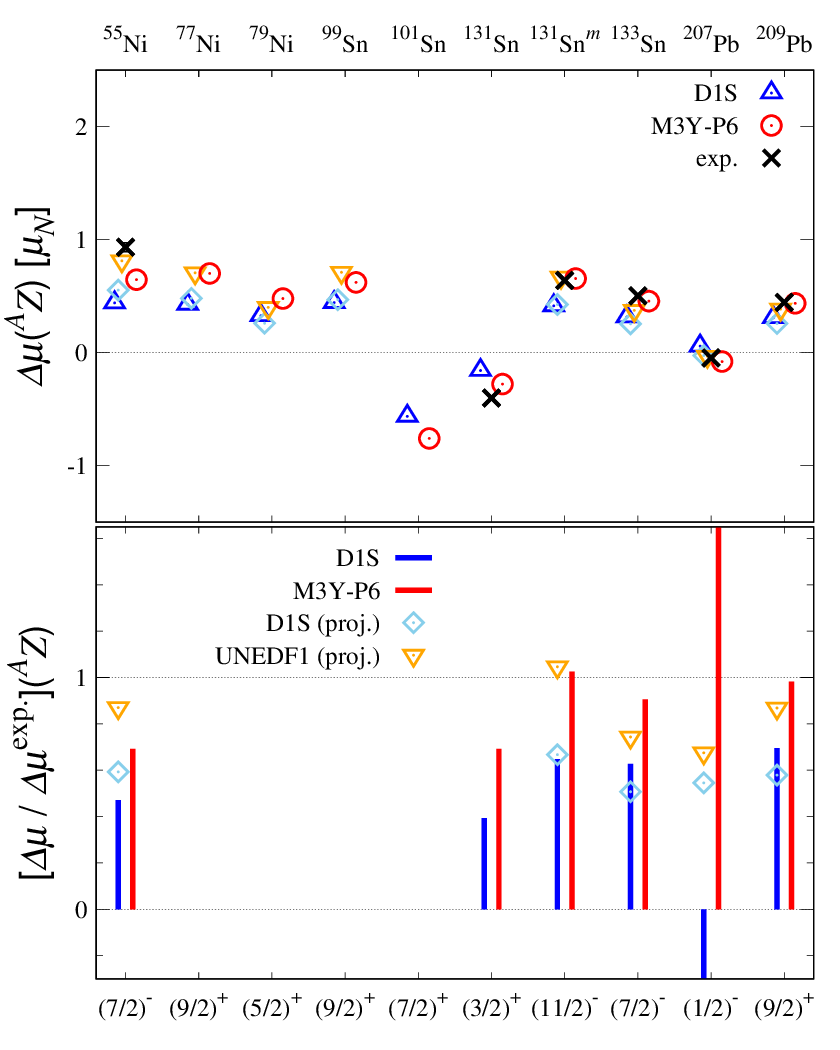}
  \caption{Upper panel:
    $\varDelta\mu$ values in odd-$N$ nuclei adjacent to the $jj$-closed shell.
    See Fig.~\protect\ref{fig:mu_ls-ls_A-odd} for other conventions.
    Experimental data are taken
    from Refs.~\protect\cite{ref:INDC-0794,ref:INDC-0816}
    and \protect\cite{ref:RBBB20} (for $^{133}$Sn).
    Lower panel:
    Calculated $\varDelta\mu$ values relative to $\varDelta\mu^\mathrm{exp.}$
    when $\mu^\mathrm{exp.}$ is available.
    See Fig.~\protect\ref{fig:mu_ls-jj_A-odd} for conventions.
\label{fig:mu_jj-jj_N-odd}}
\end{figure}

\begin{figure}
  \includegraphics[scale=0.7]{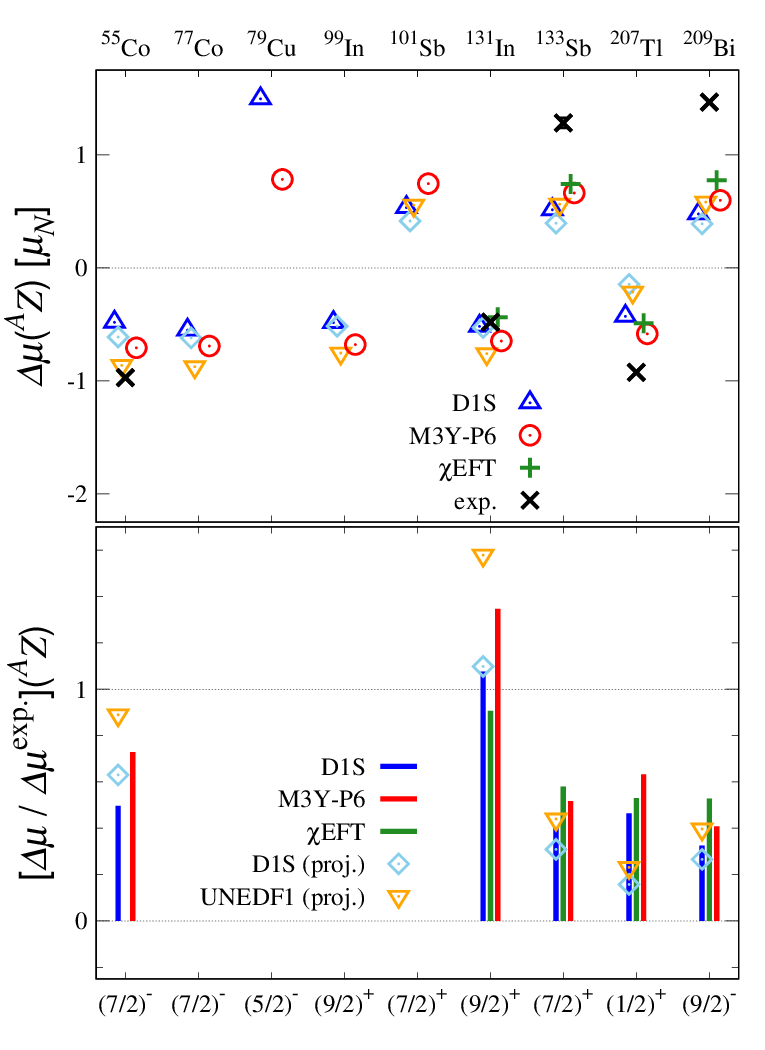}
  \caption{Upper panel:
    $\varDelta\mu$ values in odd-$Z$ nuclei adjacent to the $jj$-closed shell.
    See Fig.~\protect\ref{fig:mu_ls-ls_A-odd} for conventions.
    Experimental data are taken
    from Refs.~\protect\cite{ref:INDC-0794,ref:INDC-0816}
    and \protect\cite{ref:VGMB22} (for $^{131}$In).
    Lower panel:
    Calculated $\varDelta\mu$ values relative to $\varDelta\mu^\mathrm{exp.}$.
    Green bars represent the $\chi$EFT results
    of Ref.~\protect\cite{ref:MCSB24}.
    See Fig.~\protect\ref{fig:mu_ls-jj_A-odd} for other conventions.
\label{fig:mu_jj-jj_Z-odd}}
\end{figure}

In Fig.~\ref{fig:mu_ls-jj_A-odd},
we plot the $\varDelta\mu$ values in neighbors of the doubly-closed nuclei,
in which one of $Z$ and $N$ is an $\ell s$-closed
and the other is a $jj$-closed magic number.
It has been known that $N=14$ behaves like a magic number
in the proton-deficient region,
and $^{22}$O is analogous to a doubly-magic nucleus.
We add $^{21}$O in Fig.~\ref{fig:mu_ls-jj_A-odd}
and compare the SCMF results with the new data~\cite{ref:IGIT23}.
Figures~\ref{fig:mu_jj-jj_N-odd} and \ref{fig:mu_jj-jj_Z-odd}
display the $\varDelta\mu$ values near $jj$-closed nuclei,
separating them into odd-$N$ and odd-$Z$ ones.
At $^{131}$Sn, the data on the metastable state with $(11/2)^-$ is available,
which corresponds to the neutron single-hole state $(0h_{11/2})^{-1}$
on top of $^{132}$Sn.
It is included in Fig.~\ref{fig:mu_jj-jj_N-odd}, labeled as $^{131}$Sn$^m$.
The ratios $\varDelta\mu/\varDelta\mu^\mathrm{exp.}$ are also displayed
in Figs.~\ref{fig:mu_ls-jj_A-odd}, \ref{fig:mu_jj-jj_N-odd}
and \ref{fig:mu_jj-jj_Z-odd}.
They elucidate how well the quenching is reproduced in individual calculations.

Let us first compare two types of D1S results,
the present SCMF results (blue triangles)
and the $J$-projected results in Ref.~\cite{ref:SDBG22} (skyblue diamonds).
It is noted that the difference of the results
between Ref.~\cite{ref:SDBG22} and the present work
should be attributed to the $J$-projection
despite the difference in computational methods,
as far as both are nearly convergent.
Note that $J$-projected results have not been reported
for $^{21}$O, $(7/2)^+$ state of $^{101}$Sn, $(3/2)^+$ state of $^{131}$Sn,
and $(5/2)^-$ state of $^{79}$Cu.
We observe that the results are so close
($\mu^\mathrm{MF}\approx \mu^\mathrm{proj.}$).
It is appropriate to investigate $\mu$ via the $J$-unprojected SCMF calculation,
circumventing the doctrinal discussion mentioned in Sec.~\ref{subsec:def_M1mom}.

The SCMF approach with D1S yields correct signs of $\varDelta\mu$
(except for $^{207}$Pb),
but the degree is insufficient compared with the experimental data.
With the M3Y-P6 semi-realistic interaction,
$\varDelta\mu$ values are almost always improved from the D1S case,
though overshooting in some nuclei and remaining insufficient in others.
In the cases where D1S has already reproduced the data well, so does M3Y-P6.
The M3Y-P6 results are close to the $\chi$EFT results
for the nuclei handled in Ref.~\cite{ref:MCSB24}.
It is commented that, compared with UNEDF1
plus spin-dependent terms fitted to the data,
M3Y-P6 yields comparable or even slightly better agreement with the experiments.
The source of the dependence of $\varDelta\mu$ on the interaction
will be analyzed in the subsequent subsections.

\subsection{Electric quadrupole moments\label{subsec:results_E2mom}}

The deviation $\varDelta\mu$ reflects the CP,
\textit{i.e.}, weak erosion of the doubly-magic core.
It could be induced by quadrupole deformation, in part.
In this subsection,
we examine the quadrupole deformation predicted in the SCMF approach.
Correlations due to the spin-dependent interaction may also be important
for $\varDelta\mu$,
which will be discussed later.

\begin{figure}
  \includegraphics[scale=0.7]{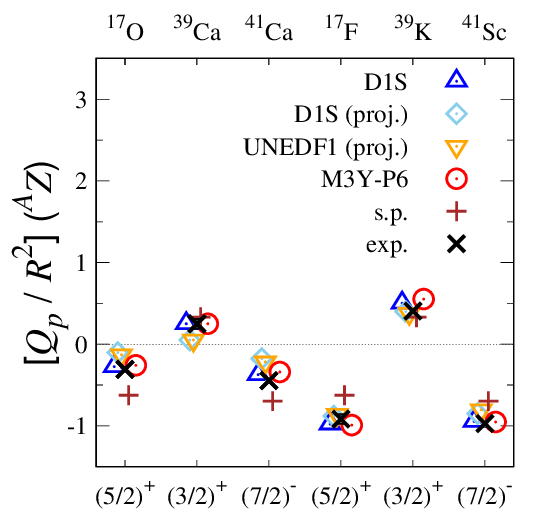}
  \caption{$Q_p/R^2$ values in nuclei of Fig.~\ref{fig:mu_ls-ls_A-odd}.
    Single-particle values ($Q_n/R^2$ for odd-$N$ nuclei)
    are shown by brown pluses.
    Other symbols are common to Fig.~\ref{fig:mu_ls-ls_A-odd}.
    Experimental data are taken from Ref.~\protect\cite{ref:Stone16}.
\label{fig:q0p_ls-ls_A-odd}}
\end{figure}

\begin{figure}
  \includegraphics[scale=0.7]{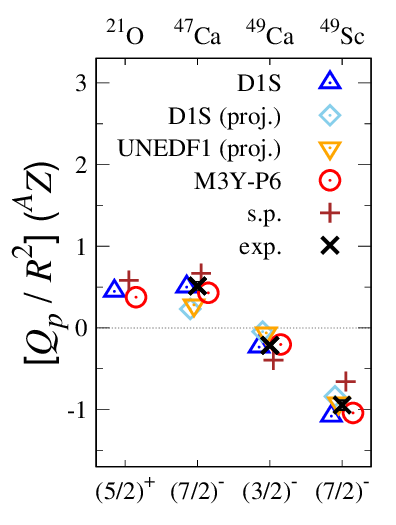}
  \caption{$Q_p/R^2$ values in nuclei of Fig.~\ref{fig:mu_ls-jj_A-odd}.
    See Fig.~\ref{fig:q0p_ls-ls_A-odd} for conventions.
    Experimental data are taken from Refs.~\protect\cite{ref:Stone16},
    \cite{ref:GBFH15} (for $^{47,49}$Ca) and \cite{ref:BKHY22} (for $^{49}$Sc).
\label{fig:q0p_ls-jj_A-odd}}
\end{figure}

\begin{figure}
  \includegraphics[scale=0.7]{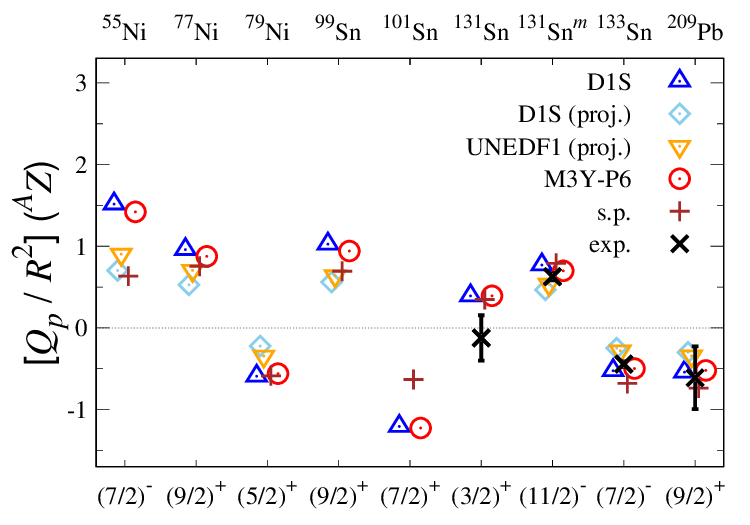}
  \caption{$Q_p/R^2$ values in nuclei of Fig.~\ref{fig:mu_jj-jj_N-odd}
    ($Q_n/R^2$ for s.p. values).
    See Fig.~\ref{fig:q0p_ls-ls_A-odd} for conventions.
    Experimental data are taken from Refs.~\protect\cite{ref:Stone16},
    \protect\cite{ref:YRBB20} (for $^{131}$Sn$^m$)
    and \protect\cite{ref:RBBB20} (for $^{133}$Sn).
\label{fig:q0p_jj-jj_N-odd}}
\end{figure}

\begin{figure}
  \includegraphics[scale=0.7]{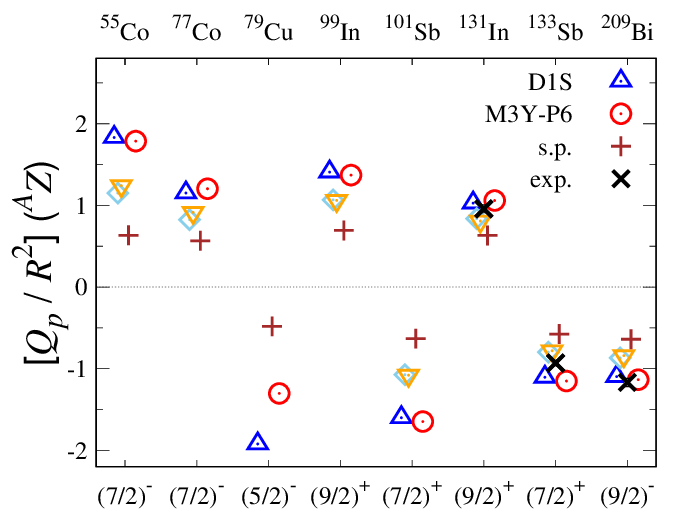}
  \caption{$Q_p/R^2$ values in nuclei of Fig.~\ref{fig:mu_jj-jj_Z-odd}.
    See Fig.~\ref{fig:q0p_ls-ls_A-odd} for conventions.
    Experimental data are taken from Refs.~\protect\cite{ref:Stone16}
    and \cite{ref:LXBB21} (for $^{133}$Sb).
\label{fig:q0p_jj-jj_Z-odd}}
\end{figure}

The $Q_p$ values in the nuclei near the doubly-magic core
are depicted in Figs.~\ref{fig:q0p_ls-ls_A-odd}, \ref{fig:q0p_ls-jj_A-odd},
  \ref{fig:q0p_jj-jj_N-odd} and \ref{fig:q0p_jj-jj_Z-odd}.
In practice, we plot $Q_p$ divided by $R^2=(1.12\,A^{1/3}\,\mathrm{fm})^2$,
to reduce dependence on the nuclear size.
As seen in Fig.~\ref{fig:q0p_ls-ls_A-odd},
the $Q_p$ values for $^{17}$F, $^{39}$K and $^{41}$Sc,
which are odd-$Z$ nuclei neighboring $^{16}$O and $^{40}$Ca,
barely deviate from the s.p. values
in both the experimental data and the SCMF results.
Consistent with $\varDelta\mu$ in Fig.~\ref{fig:mu_ls-ls_A-odd},
this outcome suggests that the erosion of the magic core
is not so strong as it is far beyond the perturbative regime,
though not negligible.
Note that $Q_n$'s are presented as the s.p. values for the odd-$N$ nuclei
$^{17}$O and $^{39,41}$Ca, instead of $Q_p$'s.
The measured and calculated $Q_p$'s are not vanishing, indicating CP,
though their absolute values are smaller than the s.p. values of $Q_n$.
It is confirmed that the SCMF calculations well reproduce
the measured $Q_p$ values both in the odd-$Z$ and odd-$N$ nuclei.

In the odd-$N$ nuclei shown in Fig.~\ref{fig:q0p_ls-jj_A-odd},
$^{21}$O, $^{47,49}$Ca,
the SCMF results of $Q_p$ reveal CP effects,
coincidentally close to the s.p. values of $Q_n$.
A similar coincidence is found for several nuclei
in Fig.~\ref{fig:q0p_jj-jj_N-odd}.
The $Q_p$ value is enhanced from the s.p. value in the odd-$Z$ nuclei,
$^{49}$Sc in Fig.~\ref{fig:q0p_ls-jj_A-odd}
and the nuclei in Fig.~\ref{fig:q0p_jj-jj_Z-odd}.
We find good agreement of the SCMF results with the available data.
The present SCMF results with D1S and M3Y-P6 are close to each other.
Thus, it is reasonable to conclude
that the SCMF calculations well describe the weak quadrupole deformation
of these nuclei,
and the degree of deformation is not sensitive to the effective interactions.

\subsection{Occupation numbers on individual $(\ell  j)$ component
  \label{subsec:occ}}

We next investigate occupation numbers on individual $(\ell j)$ components,
$\langle N_{\tau,\ell j}\rangle$.
At the spherical limit,
the occupation numbers are determined
by those of the adjacent doubly-magic core and the last nucleon,
and we denote them by $N_{\tau,\ell j}^\mathrm{sph}$.
Take neutron orbitals at $^{41}$Ca as an example.
We have $N_{n,s_{1/2}}^\mathrm{sph}=4$
because of the occupation of the $0s_{1/2}$ and $1s_{1/2}$ orbits,
$N_{n,p_{3/2}}^\mathrm{sph}=4$, $N_{n,p_{1/2}}^\mathrm{sph}=2$,
$N_{n,d_{5/2}}^\mathrm{sph}=6$, $N_{n,d_{3/2}}^\mathrm{sph}=4$,
and $N_{n,f_{7/2}}^\mathrm{sph}=1$ because of the last neutron.
Instead of the occupation numbers themselves,
their difference from those at the spherical limit,
$\varDelta\langle N_{\tau,\ell j}\rangle
:= \langle N_{\tau,\ell j}\rangle - N_{\tau,\ell j}^\mathrm{sph}$,
will be plotted.
$|\varDelta\langle N_{\tau,\ell j}\rangle|\ll 1$ verifies
that the breaking of magicity stays weak.

\begin{figure}
  \includegraphics[scale=0.7]{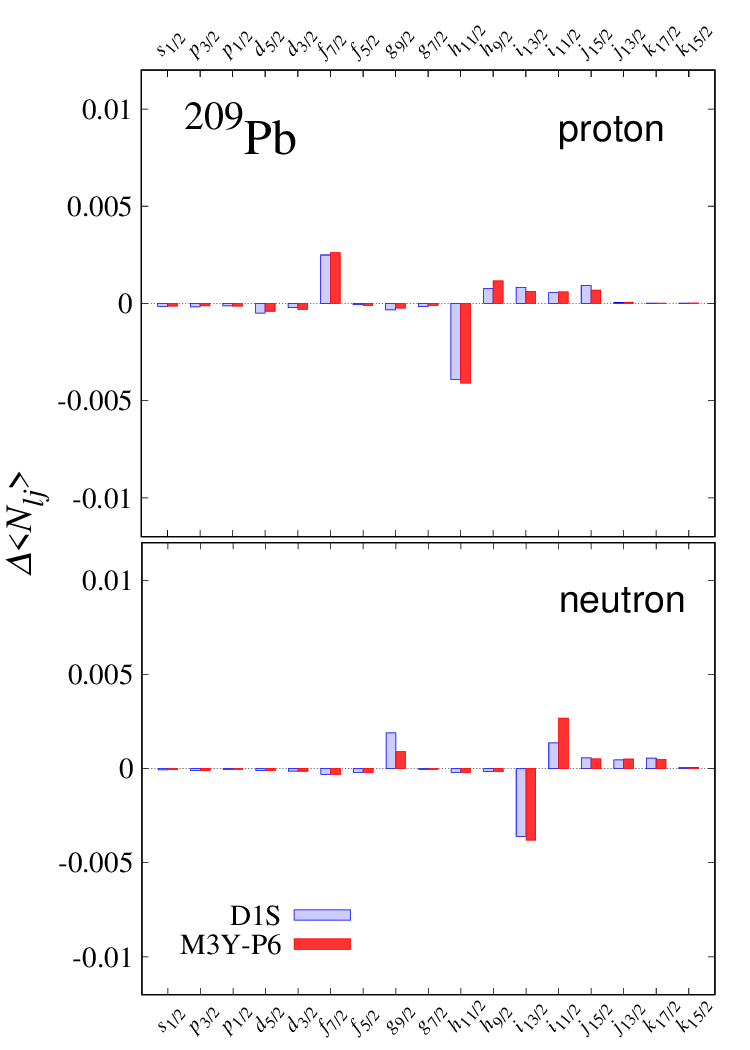}
  \caption{$\varDelta\langle N_{\tau,\ell j}\rangle$ at $^{209}$Pb.
    The SCMF results with M3Y-P6 and D1S
    are shown by red and blue bars, respectively.
\label{fig:occlj_Pb209}}
\end{figure}

\begin{figure}
  \includegraphics[scale=0.7]{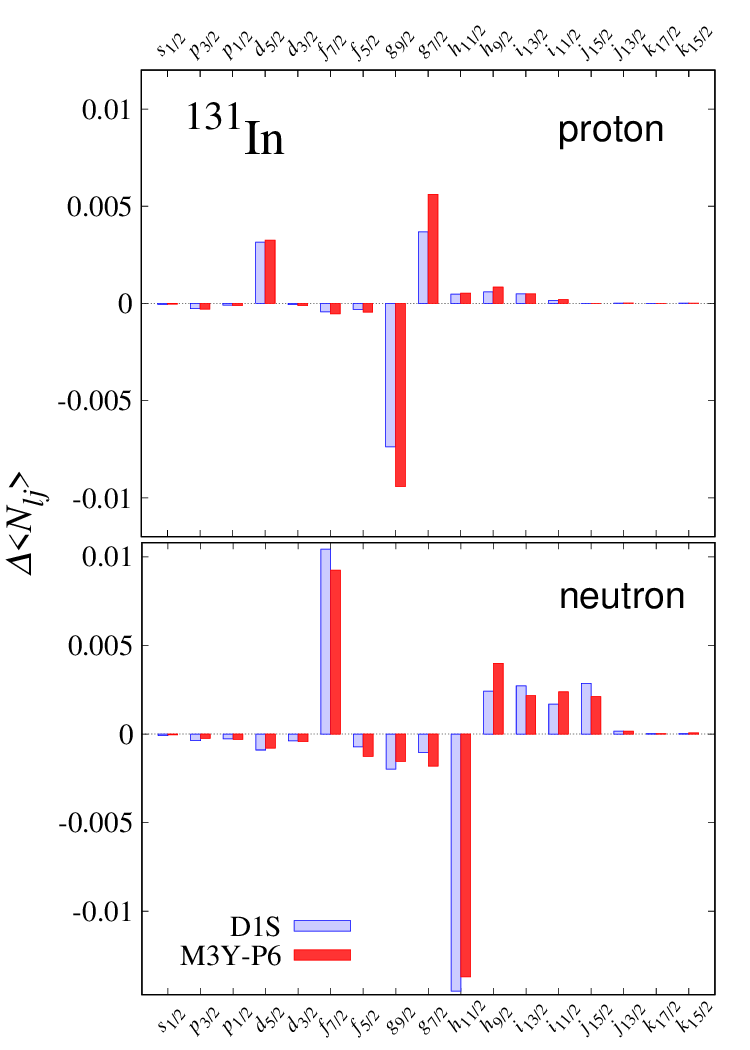}
  \caption{$\varDelta\langle N_{\tau,\ell j}\rangle$ at $^{131}$In.
    Conventions are common with Fig.~\protect\ref{fig:occlj_Pb209}.
\label{fig:occlj_In131}}
\end{figure}

Figures~\ref{fig:occlj_Pb209} and \ref{fig:occlj_In131}
depict $\varDelta\langle N_{\tau,\ell j}\rangle$ at $^{209}$Pb and $^{131}$In,
respectively.
We do not find a notable qualitative difference in other nuclei.
At $^{209}$Pb, the last neutron occupies the $1g_{9/2}$ orbit
at the spherical limit.
We observe that it induces proton excitation from $h_{11/2}$ to $f_{7/2}$
both in the D1S and M3Y-P6 results.
As these two orbits strongly couple under the quadrupole field,
they imply the relevance of the quadrupole deformation,
even though it is weak.
Similar excitations relevant to quadrupole deformation are found
on the neutron side ($i_{13/2}$ to $g_{9/2}$)
and at $^{131}$In ($g_{9/2}$ to $d_{5/2}$ on the proton side
and $h_{11/2}$ to $f_{7/2}$ on the neutron side).
In addition, we find sizable excitations
to the neutron $i_{11/2}$ component at $^{209}$Pb
and the proton $g_{7/2}$ component at $^{131}$In.
They are $\ell s$ partners of the high-$j$ occupied orbits
of the doubly-magic cores,
and the excitations exhibit correlation with respect to the spin d.o.f.
Though weaker,
excitations to the $\ell s$ partners are visible
also on the proton (neutron) side at $^{209}$Pb ($^{131}$In).
As discussed in Subsec.~\ref{subsec:perturb},
these excitations trigger $\varDelta\mu$
within the lowest-order perturbation.

It is noticed that excitations to the $\ell s$ partners
are always stronger in the M3Y-P6 results than in the D1S results,
implying strong spin correlations in M3Y-P6.
This difference in $\varDelta\langle N_{\tau,\ell j}\rangle$
accounts for the interaction-dependence of $\varDelta\mu$.
It will be instructive to specify
which channel of the interaction gives rise to this difference.

\subsection{Analysis via lowest-order perturbation\label{subsec:perturb}}

As long as the erosion of the doubly-magic core is not serious,
analysis in terms of the perturbation theory should be useful.
Owing to its linear nature,
the lowest-order perturbation allows us
to separate the contributions of each channel of the interaction.

\begin{figure}
  \includegraphics[scale=1.0]{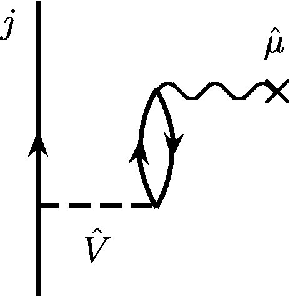}
  \caption{Goldstone diagram for the lowest-order contribution
    $\varDelta\mu^\mathrm{pert.}$.
    $\hat{V}$ represents the interaction
    and $\hat{\mu}$ the $M1$ operator.
\label{fig:diagram}}
\end{figure}

In the lowest-order perturbation,
$\varDelta\mu$ is expressed by the diagram of Fig.~\ref{fig:diagram}
and the following equation~\cite{ref:AH54},
\begin{equation}
  \varDelta\mu^\mathrm{pert.}(^A Z)
  = -2\,\big\langle j\,m=j\big|\mu^{(1)}_0
  \big|j\,(j_2\,j_1^{-1})^{(1)};j\,m=j\big\rangle\,
  \frac{\big\langle j\,(j_2\,j_1^{-1})^{(1)};j\big|V\big|j\big\rangle}
       {\epsilon_{j_2}-\epsilon_{j_1}}\,.
\label{eq:dmu_pert}
\end{equation}
Here $|j\rangle$ is the unperturbed state,
which has a single particle or hole on the orbit $j$
on top of the doubly-magic core,
and $\big|j\,(j_2\,j_1^{-1})^{(1)};j\big\rangle$
is composed of the particle or hole on $j$
and the CP component $(j_2\,j_1^{-1})^{(1)}$, the excitation from $j_1$ to $j_2$.
For the operator $\mu^{(1)}$ to act on,
the polarization is restricted to the $1^+$ spin-parity,
and $\varDelta\mu^\mathrm{pert.}$ of Eq.~\eqref{eq:dmu_pert} thereby stands for
CP due to the spin correlation.
The coupling of the particle (or hole) on $j$ to $(j_2\,j_1^{-1})^{(1)}$
must result in $J=j$ due to the angular-momentum conservation via $V$.
All the relevant s.p. energies ($\epsilon_{j_1}$, $\epsilon_{j_2}$)
and functions are obtained by the SCMF calculations at the doubly-magic nucleus.
Note that
$\big\langle j\,m\big|\mu^{(1)}_0\big|j\,(j_2\,j_1^{-1})^{(1)};j\,m\big\rangle$
is evaluated from the s.p. matrix element $\langle j_1||\mu^{(1)}||j_2\rangle$,
and $\big\langle j\,(j_2\,j_1^{-1})^{(1)};j\big|V\big|j\big\rangle$
from the two-body elements of the interaction
$\langle j\,j_2|V|j\,j_1\rangle$.

The perturbative picture well accounts for
the reason why $\varDelta\mu$ is small near the $\ell s$-closed nuclei,
as shown in Fig.~\ref{fig:mu_ls-ls_A-odd}.
Because the $\mu^{(1)}$ operator in Eq.~\eqref{eq:mm-op} does not change
the orbital angular momentum of the s.p. state,
$j_1$ and $j_2$ in the polarization component $(j_2\,j_1^{-1})^{(1)}$
should be an $\ell s$ partner.
The lowest-order term $\varDelta\mu^\mathrm{pert.}$ in Eq.~\eqref{eq:dmu_pert}
requires occupied $j_1$ and unoccupied $j_2$,
which is forbidden at the $\ell s$-closed core.
The smallness of measured $\varDelta\mu$ around the $\ell s$-closed nuclei
is evidence for the dominance of CP in $\varDelta\mu$.

The SCMF results $\varDelta\mu^\mathrm{MF}$ exhibit
effects beyond $\varDelta\mu^\mathrm{pert.}$.
For instance, the quadrupole polarization like $[(p1f_{7/2})(p0h_{11/2}^{-1})]$
observed in Fig.~\ref{fig:occlj_Pb209}
is distinguished from the $(j_2\,j_1^{-1})^{(1)}$ polarization.
As discussed in Subsecs.~\ref{subsec:results_E2mom} and \ref{subsec:occ},
M3Y-P6 and D1S are similar in the quadrupole correlation
but have difference in the spin correlation.
We investigate through $\varDelta\mu^\mathrm{pert.}$
which part of the interaction makes this difference.

\begin{figure}
  \includegraphics[scale=0.7]{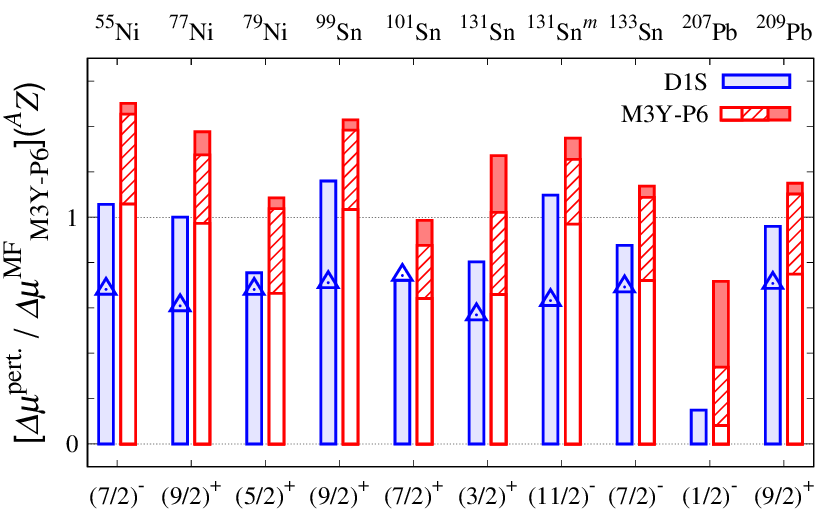}
  \caption{$\varDelta\mu^\mathrm{pert.}$ relative to $\varDelta\mu^\mathrm{MF}$
    with M3Y-P6 for nuclei of Fig.~\ref{fig:mu_jj-jj_N-odd}.
    Red (blue) bars represent $\varDelta\mu^\mathrm{pert.}$ with M3Y-P6 (D1S).
    $\varDelta\mu^\mathrm{MF}$ with D1S
    (relative to $\varDelta\mu^\mathrm{MF}$ with M3Y-P6)
    is shown in blue triangles.
    In $\varDelta\mu^\mathrm{pert.}$ with M3Y-P6,
    hatched and filled areas display
    the contribution of $V^{(\mathrm{C})}_\mathrm{OPEP}$ and $V^{(\mathrm{TN})}$.
\label{fig:pert-mu_jj-jj_N-odd}}
\end{figure}

The ratio $\varDelta\mu^\mathrm{pert.}/\varDelta\mu^\mathrm{MF}$
will provide a measure
of how well $\varDelta\mu^\mathrm{pert.}$ approximates $\varDelta\mu$.
We adopt the $\varDelta\mu^\mathrm{MF}$ values with M3Y-P6 as a reference.
In Fig.~\ref{fig:pert-mu_jj-jj_N-odd},
$\varDelta\mu^\mathrm{pert.}$ relative to $\varDelta\mu^\mathrm{MF}$ with M3Y-P6
is depicted for the nuclei handled in Fig.~\ref{fig:mu_jj-jj_N-odd}.
$\varDelta\mu^\mathrm{pert.}$ with D1S is also presented
in terms of the ratio to $\varDelta\mu^\mathrm{MF}$ with M3Y-P6.
Deviation of $\varDelta\mu^\mathrm{pert.}/\varDelta\mu^\mathrm{MF}$ from unity
shows that the perturbation of Eq.~\eqref{eq:dmu_pert} is not sufficient
for a fully quantitative evaluation of $\varDelta\mu$.
It also holds for the results with D1S,
as recognized by comparing the blue bars and triangles
in Fig.~\ref{fig:pert-mu_jj-jj_N-odd}.
Still, the deviation of $\varDelta\mu^\mathrm{pert.}$
from $\varDelta\mu^\mathrm{MF}$ stays within a factor of two,
and analysis through $\varDelta\mu^\mathrm{pert.}$ is useful
for the present purpose.

Contributions of $V^{(\mathrm{C})}_\mathrm{OPEP}$ and $V^{(\mathrm{TN})}$
are assessed by inserting them into $V$ in Eq.~\eqref{eq:dmu_pert},
and shown in Fig.~\ref{fig:pert-mu_jj-jj_N-odd}.
Although the effective interaction of the SCMF calculations
affects the s.p. energies $\epsilon_j$
as well as $\langle V\rangle$ in Eq.~\eqref{eq:dmu_pert},
it is more or less tuned so as to give reasonable s.p. energies.
We neglect the influence of the individual channel on $\epsilon_j$,
keeping $\epsilon_j$ obtained from the full SCMF calculation.
It is clearly seen that $V^{(\mathrm{C})}_\mathrm{OPEP}$ substantially contributes
to $\varDelta\mu$,
as expected from its effects on the spin properties discussed
in Sec.~\ref{subsec:prop_effint}.
Whereas $V^{(\mathrm{TN})}$ affects $\varDelta\mu$
cooperatively with $V^{(\mathrm{C})}_\mathrm{OPEP}$,
its effects are smaller than those of $V^{(\mathrm{C})}_\mathrm{OPEP}$;
sizable for some nuclei but not so for others.
Concerning the tensor force,
it is fair to remind that $V^{(\mathrm{TN})}$ affects the s.p. energies.
While some of the effects may be incorporated into the central channels,
the explicit inclusion of $V^{(\mathrm{TN})}$ is important
in their $Z$- and $N$-dependence.

Intriguingly, the summed contributions
of $V^{(\mathrm{C})}_\mathrm{OPEP}$ and $V^{(\mathrm{TN})}$
are close to the difference of $\varDelta\mu^\mathrm{pert.}$
between M3Y-P6 and D1S.
While the spin-dependent channels of the effective interaction
have been difficult to fix from a fully phenomenological standpoint,
realistic spin-isospin and tensor channels
connected to the bare nucleonic interaction,
particularly the channel originating from the one-pion exchange,
are helpful.
The importance of the pion exchange is harmonious with the $\chi$EFT picture.

\section{Summary and discussions\label{sec:summary}}

We have investigated magnetic dipole ($M1$) moments
adjacent to doubly-magic nuclei
using self-consistent mean-field (SCMF) approaches.
Hartree-Fock calculations assuming the $J_z$ (axial), $\mathcal{P}$ (parity)
and $\mathcal{RT}$
($\mathcal{R}$ is the reflection with respect to the $y$-axis) symmetries,
but allowing the breaking of the $\mathcal{T}$ (time-reversal) symmetry,
have been implemented with finite-range interactions.
No adjustable parameter is newly introduced.
The M3Y-P6 interaction, which keeps realistic spin-isospin and tensor channels,
is primarily applied,
and its results are compared with those obtained from other interactions.
The $M1$ moments deviate from the Schmidt values,
implying the quenching of the spin matrix elements.
The source of the deviation in the SCMF results
has been analyzed via the quadrupole moments, occupation numbers,
and comparison with the results of the lowest-order perturbation.

The SCMF results with M3Y-P6 are in reasonable agreement
with the measured $M1$ moments of nuclei neighboring the doubly-magic core.
The deviation from the Schmidt values is well described
for the nuclei adjacent to $jj$-closed magicity,
owing to the cooperative effects of weak quadrupole deformation
and spin correlation.
Compared with the results of D1S,
which is one of the standard SCMF effective interactions,
M3Y-P6 yields improvement on almost all nuclei under investigation.
The M3Y-P6 results are comparable to, or even slightly better than,
the UNEDF1 results of Ref.~\cite{ref:SDBG22},
whose spin-isospin channel was fitted to the measured $M1$ moments.
The SCMF results with M3Y-P6 are also close to the shell-model results
with the interaction derived from the chiral effective-field theory ($\chi$EFT)
in Ref.~\cite{ref:MCSB24}.
In comparison with the D1S results,
the improvement can be attributed primarily to the spin-isospin channel
and secondarily to the tensor channel in M3Y-P6,
which are considered realistic and retain a microscopic origin.
The UNEDF1 results suggest that
a central spin-isospin channel can express certain parts of these effects,
even imitating some tensor-force effects.
It is still noted that the distinction between the central and tensor channels
is vital when extrapolating the properties in an astrophysical environment.

Further consequences can be addressed.
Good agreement of the D1S results without the $J$-projection
and those with the projection in Ref.~\cite{ref:SDBG22}
suggests that the SCMF wave-functions may be used to describe $M1$ moments
without handling additional rotational correlations.
This implies that the $M1$ moments can be a principal variable
of the generalized Kohn-Sham (KS) framework.
In the results from $\chi$EFT~\cite{ref:MCSB24},
the meson-exchange current (MEC) has improved the agreement
with experimental data,
which counteracts the CP in most cases~\cite{ref:Tow87,ref:ASBH88,ref:MCSB24}.
On the other hand, the coupling of the virtual isobar excitation
to the magnetic field has not been taken into account.
Since these effects tend to cancel each other~\cite{ref:Tow87,ref:ASBH88},
though not completely,
the $M1$ moments could be calculated
by the one-body operator of Eq.~\eqref{eq:mm-op} to moderate precision.
The results reported in Refs.~\cite{ref:BE17,ref:PHGM21,ref:VGMB22,ref:BDDK23}
indicate that the KS approach is promising to describe $M1$ moments in nuclei
departing from the doubly-magic core.
However, it is not easy to completely fix the spin-dependent channels
of the effective interaction by fitting.
The present work exemplifies
that microscopic theory could provide good guidance.
Whereas the one-pion exchange is an important ingredient
of the central spin-isospin channel,
there are few codes applicable to the Yukawa interaction,
with the exception of the code
used in the present work~\cite{ref:NS02,ref:Nak06,ref:Nak08}.
Along this line,
local approximation based on the density-matrix expansion
(DME)~\cite{ref:NV72,ref:CB78,ref:DCK10}
has been applied to the Fock term in, \textit{e.g.}, Ref.~\cite{ref:ZBFN24}.
According to Ref.~\cite{ref:ZCBF21},
the DME leads to errors of a few to ten percent,
which is moderately good but not excellent precision.
It is desirable to keep paying attention to the precision of the DME,
particularly for the spin-dependent channels~\cite{ref:DCK10}.

Within the perturbation theory,
the $M1$ moments near the $\ell s$-closed nuclei
hardly deviate from the Schmidt values at the lowest order,
although the higher-order terms
are not necessarily negligible~\cite{ref:Tow87,ref:ASBH88,ref:SIA74}.
While the present SCMF results are consistent with the weak quenching
in vicinity of the $\ell s$-closure,
they fail to describe the measured $\Delta\mu$ values precisely,
as seen for $^{41}$Ca, $^{39}$K and $^{41}$Sc in Fig.~\ref{fig:mu_ls-ls_A-odd}.
In vicinity of the $jj$-closure,
apparent discrepancy exists at the $Z=\mathrm{odd}$ nuclei
$^{133}$Sb, $^{207}$Ti and $^{209}$Bi,
while agreement is remarkable at $^{131}$Sn$^m$, $^{133}$Sn and $^{209}$Pb.
We note that the $\chi$EFT results also deviate from the data
at the above $Z=\mathrm{odd}$ nuclei~\cite{ref:MCSB24}.
At $^{209}$Bi,
it has been shown that the MEC could exceptionally facilitate
quenching~\cite{ref:ASBH88,ref:LM18,ref:MCSB24}.
It would be of interest to incorporate
corrections due to the MEC,
which are represented by two-body operators~\cite{ref:BHGW91,ref:SHMB23},
and isobar excitations
into the SCMF calculations of $M1$ moments at some nuclei,
\textit{e.g.}, $^{209}$Bi,
though it is beyond the scope of the present work.

\appendix

\section{Relation of time-reversality to other symmetries in one-body fields
  \label{app:T-sym}}

Symmetries concerning the rotation may restrict
the effect of the time-reversality (\textit{i.e.}, $\mathcal{T}$-symmetry).
In this Appendix,
we discuss the relations of the $\mathcal{T}$-symmetry
to the spherical ($\mathbf{J}$) or axial ($J_z$), parity ($\mathcal{P}$),
and reflection ($\mathcal{R}$) symmetries
in terms of the one-body density matrix,
particularly in the spherical-basis representation.
The argument in this Appendix serves as one of the grounds
for the numerical calculations in the present work.
The self-consistent symmetries were minutely investigated
for the local densities and currents in Ref.~\cite{ref:RDN10}.
In contrast, our discussion is not constrained to local fields
and may be transparent in specific cases.

The mean fields are represented
by the one-body density matrix
$\varrho_{kk'}\,(:=\langle\Phi|c_{k'}^\dagger c_k|\Phi\rangle)$
and the pairing tensor
$\kappa_{kk'}\,(:=\langle\Phi|c_{k'} c_k|\Phi\rangle)$, in general.
Here, $k$ and $k'$ denote the s.p. bases,
and $c_k^\dagger$ ($c_k$) is
the corresponding creation (annihilation) operator.
Symmetries in the mean fields are nothing but
those in $\varrho_{kk'}$ and $\kappa_{kk'}$.
Note that $\varrho_{kk'}$ is hermitian
(\textit{i.e.}, $\varrho_{k'k}=\varrho_{kk'}^\ast$)
and $\kappa_{kk'}$ is skew-symmetric
(\textit{i.e.}, $\kappa_{k'k}=-\kappa_{kk'}$).
We restrict ourselves here to the symmetries in $\varrho_{kk'}$;
those in $\kappa_{kk'}$ can be discussed analogously.
Adopting the spherical bases, we take $k=(\nu\ell jm)$,
where $\ell$, $j$ and $m$ are the orbital angular momentum,
the summed angular momentum and its $z$-component, respectively.
The label $\nu$ distinguishes the radial wave functions.
We do not consider proton-neutron mixing in the s.p. states,
and omit the isospin index for the sake of simplicity.
By using the s.p. basis function $\phi_k(\mathbf{r}\sigma)$,
where $\sigma$ is the spin index,
$\varrho_{kk'}$ is readily transformed to the coordinate representation
$\varrho(\mathbf{r}\sigma,\mathbf{r}'\sigma')$,
\begin{equation}
  \varrho_{kk'} = \sum_{\sigma\sigma'} \int d^3r\,d^3r'\,
  \varrho(\mathbf{r}\sigma,\mathbf{r}'\sigma')\,
  \phi_k^\ast(\mathbf{r}\sigma)\,\phi_{k'}(\mathbf{r}'\sigma')\,.
\label{eq:rho-trans}\end{equation}

The time-reversed state of $k$ is denoted by $\bar{k}$.
In the concrete, $\bar{k}$ is $(\nu\ell j\,-m)$ with an appropriate phase,
for which we use the convention $(-)^{j+\ell-m}$.
Namely, the s.p. bases are $\mathcal{RPT}$-invariant.
The $\mathcal{T}$-symmetry in the mean-fields is recognized as,
\begin{equation}
  \mbox{$\mathcal{T}$-symmetry\,:}\quad
  \varrho_{\bar{k}\bar{k'}} = \varrho_{kk'}^\ast \,.
\label{eq:rho-T}\end{equation}
The complex conjugate on the rhs is a result of the $\mathcal{T}$ operation
on the transformation coefficients of the s.p. states
[\textit{e.g.}, $U$ and $V$ in Eq.~\eqref{eq:Jz-Bogtr}].

Under the $\mathbf{J}$-symmetry,
$\varrho_{kk'}$ is diagonal with respect to $j$ and $m$,
and has no dependence on $m$,
\begin{equation}
  \varrho_{kk'} = \delta_{jj'}\delta_{mm'}\,\varrho^{(j)}_{\nu\ell,\nu'\ell'}\,.
\label{eq:rho-J}\end{equation}
The $\mathcal{R}$-symmetry plays no additional roles in this case.
Equation~\eqref{eq:rho-T} is reduced to,
applying the present phase convention,
\begin{equation}
  \mbox{$(\mathbf{J}+\mathcal{T})$-symmetry\,:}\quad
  (-)^{\ell-\ell'}\,\varrho^{(j)}_{\nu\ell,\nu'\ell'}
  = \varrho^{(j)^\ast}_{\nu\ell,\nu'\ell'} \,.
\label{eq:rho-J+T}\end{equation}
We usually have $\mathcal{P}$-symmetry when we maintain spherical symmetry.
Then, $\varrho_{kk'}$ is also diagonal with respect to $\ell$,
and Eq.~\eqref{eq:rho-J} reads,
\begin{equation}
  \varrho_{kk'} = \delta_{jj'}\delta_{\ell\ell'}\delta_{mm'}\,
  \varrho^{(\ell j)}_{\nu,\nu'}\,.
  \label{eq:rho-J+P}\end{equation}
The condition of the $\mathcal{T}$-symmetry \eqref{eq:rho-T} becomes
\begin{equation}
  \mbox{$(\mathbf{J}+\mathcal{P}+\mathcal{T})$-symmetry\,:}\quad
  \varrho^{(\ell j)}_{\nu,\nu'} = \varrho^{(\ell j)^\ast}_{\nu,\nu'} \,.
\label{eq:rho-J+P+T}\end{equation}
Equation~\eqref{eq:rho-J+P+T} indicates,
since $\nu$ is the label for the radial part of the s.p. basis,
that the violation of the $\mathcal{T}$-symmetry may take place
only through the off-diagonal element with respect to the radial part,
irrelevant to the angular-spin part.
It is not a surprising consequence
since the $(\mathbf{J}+\mathcal{P})$-symmetry
fixes the angular-spin part of the s.p. states.

Under the $J_z$-symmetry,
$\varrho_{kk'}$ is diagonal with respect to $m$ but not necessarily to $j$,
\begin{equation}
  \varrho_{kk'} = \delta_{mm'}\,\varrho^{(m)}_{\nu\ell j,\nu'\ell'j'}\,.
\label{eq:rho-Jz}\end{equation}
Equation~\eqref{eq:rho-T} is reduced to,
\begin{equation}
  \mbox{$(J_z+\mathcal{T})$-symmetry\,:}\quad
  (-)^{j+\ell-j'-\ell'}\,\varrho^{(-m)}_{\nu\ell j,\nu'\ell'j'}
  = \varrho^{(m)^\ast}_{\nu\ell j,\nu'\ell'j'} \,.
\label{eq:rho-Jz+T}\end{equation}
If we have the $\mathcal{P}$-symmetry in addition,
Eq.~\eqref{eq:rho-Jz+T} leads to
\begin{equation}
  \mbox{$(J_z+\mathcal{P}+\mathcal{T})$-symmetry\,:}\quad
 (-)^{j-j'}\,\varrho^{(-m)}_{\nu\ell j,\nu'\ell'j'}
  = \varrho^{(m)^\ast}_{\nu\ell j,\nu'\ell'j'} \,.
\label{eq:rho-Jz+P+T}\end{equation}
When we have the $\mathcal{R}$-symmetry,
even without the $\mathcal{P}$-symmetry,
the density matrix in Eq.~\eqref{eq:rho-Jz} should satisfy,
\begin{equation}
  \mbox{$(J_z+\mathcal{R})$-symmetry\,:}\quad
  (-)^{j-j'}\,\varrho^{(-m)}_{\nu\ell j,\nu'\ell'j'}
  = \varrho^{(m)}_{\nu\ell j,\nu'\ell'j'} \,.
\label{eq:rho-Jz+R}\end{equation}
The $\mathcal{RT}$-symmetry
rather than the individual $\mathcal{R}$ or $\mathcal{T}$ symmetry
yields
\begin{equation}
  \mbox{$(J_z+\mathcal{RT})$-symmetry\,:}\quad
  (-)^{\ell-\ell'}\,\varrho^{(m)}_{\nu\ell j,\nu'\ell'j'}
  = \varrho^{(m)^\ast}_{\nu\ell j,\nu'\ell'j'} \,,
\label{eq:rho-Jz+RT}\end{equation}
and the $\mathcal{RPT}$-symmetry derives
\begin{equation}
  \mbox{$(J_z+\mathcal{RPT})$-symmetry\,:}\quad
 \varrho^{(m)}_{\nu\ell j,\nu'\ell'j'}
  = \varrho^{(m)\ast}_{\nu\ell j,\nu'\ell'j'} \,.
\label{eq:rho-Jz+RPT}\end{equation}

\section{Mean-field equation under $J_z$ conservation\label{app:axial}}

While the Hartree-Fock calculations have been applied in the present work,
in this Appendix we derive the mean-field equation in a more general respect;
namely, the Hartree-Fock-Bogoliubov (HFB) equation.
The creation and annihilation operators
associated with the s.p. basis $\phi_{\nu\ell jm}$
are denoted by $c_{\nu\ell jm}^\dagger$ and $c_{\nu\ell jm}$.
We define the modified annihilation operator
as $\tilde{c}_{\nu\ell jm}=(-)^{j+m} c_{\nu\ell j -m}$
as in Ref.~\cite{ref:Nak06}.
Under the $J_z$ conservation,
the Bogoliubov transformation is given by
\begin{equation}
  \alpha_{n,m}^\dagger = \sum_{\nu\ell j}
  \big[ U^{(m)}_{\nu\ell j,n} c_{\nu\ell jm}^\dagger
    + (-)^{j+m}\,V^{(-m)}_{\nu\ell j,n} \tilde{c}_{\nu\ell jm}\big]\,.
\label{eq:Jz-Bogtr} \end{equation}
The density matrix and the pairing tensor are then expressed by
\begin{equation}
  \varrho^{(m)}_{\nu\ell j,\nu'\ell'j'}
  = \sum_n V^{(-m)\ast}_{\nu\ell j,n}\,V^{(-m)}_{\nu'\ell'j',n}\,,\quad
  \kappa^{(m)}_{\nu\ell j,\nu'\ell'j'}
  = \sum_n V^{(-m)\ast}_{\nu\ell j,n}\,U^{(-m)}_{\nu'\ell'j',n}\,,
\label{eq:Jz-dnsmat} \end{equation}
having the following properties,
\begin{equation}
  \varrho^{(m)}_{\nu'\ell'j',\nu\ell j} = \varrho^{(m)\ast}_{\nu\ell j,\nu'\ell'j'}\,,\quad
  \kappa^{(-m)}_{\nu'\ell'j',\nu\ell j} = -\kappa^{(m)}_{\nu\ell j,\nu'\ell'j'}\,.
\label{eq:Jz-dnsmat-sym} \end{equation}
It is found that the HFB Hamiltonian has the structure as
\begin{equation}
  \mathcal{H} = \begin{pmatrix}
    \mathsf{h}^{(m)} & 0 & 0 & \mathsf{\Delta}^{(m)} \\
    0 & \mathsf{h}^{(-m)} & \mathsf{\Delta}^{(-m)} & 0 \\
    0 & -\mathsf{\Delta}^{(m)\ast} & -\mathsf{h}^{(m)\ast} & 0 \\
    -\mathsf{\Delta}^{(-m)} & 0 & 0 & -\mathsf{h}^{(-m)\ast} \end{pmatrix}\,,
\label{eq:Jz-HFBhamil} \end{equation}
leading to the HFB equation as
\begin{equation}\begin{split}
  \mathcal{H}'\,\begin{pmatrix} \mathsf{U}^{(m)} & \mathsf{V}^{(-m)\ast} \\
    \mathsf{V}^{(m)} & \mathsf{U}^{(-m)\ast} \end{pmatrix}
  =& \begin{pmatrix} \mathsf{U}^{(m)} & \mathsf{V}^{(-m)\ast} \\
    \mathsf{V}^{(m)} & \mathsf{U}^{(-m)\ast} \end{pmatrix}\,
  \begin{pmatrix} \mathrm{diag}[\varepsilon^{(m)}] & 0 \\
    0 & -\mathrm{diag}[\varepsilon^{(-m)}] \end{pmatrix}\,;\\
  &\qquad \mathcal{H}'
  = \begin{pmatrix} \mathsf{h}^{(m)} & \mathsf{\Delta}^{(m)} \\
    -\mathsf{\Delta}^{(-m)\ast} & -\mathsf{h}^{(-m)\ast} \end{pmatrix}\,.
\end{split}\label{eq:Jz-HFBeq} \end{equation}
From the total energy of the system $E$
represented by $(\varrho,\kappa,\kappa^\ast)$,
$\mathsf{h}^{(m)}$ is defined by the derivative of $E$
with respect to $\varrho$,
and $\mathsf{\Delta}^{(m)}$ by the derivative with respect to $\kappa^\ast$.
Corresponding to Eq.~\eqref{eq:Jz-dnsmat-sym},
we have $\mathsf{h}^{(m)T}=\mathsf{h}^{(m)\ast}$
and $\mathsf{\Delta}^{(-m)T}=-\mathsf{\Delta}^{(m)}$,
assuring $\mathcal{H}'$ to be hermitian.
It is remarked that
$\mathrm{dim}(\mathcal{H}')=\frac{1}{2}\mathrm{dim}(\mathcal{H})$
and Eq.~\eqref{eq:Jz-HFBeq} yields
all the eigensolutions without doubling, even in the HFB framework.
Its origin is the property of the pairing tensor
in Eq.~\eqref{eq:Jz-dnsmat-sym},
which connects the matrices for $-m$ with those for $m$,
leading to the structure of $\mathcal{H}$ shown in Eq.~\eqref{eq:Jz-HFBhamil}.
Whereas we do not necessarily impose orthogonality
for the radial part of s.p. bases~\cite{ref:NS02,ref:Nak06,ref:Nak08},
we write down Eq.~\eqref{eq:Jz-HFBeq} for orthogonal bases.
The norm matrix should be supplemented
for non-orthogonal bases~\cite{ref:Nak06}.

When we have $\mathcal{P}$-conservation,
Eq.~\eqref{eq:Jz-HFBeq} is separable into sectors of individual parities.
Since we adopt the $\mathcal{RPT}$-invariant basis functions here,
we can restrict all the matrices $\mathsf{U}$, $\mathsf{V}$,
$\mathsf{h}$ and $\mathsf{\Delta}$ to be real numbers
under the $\mathcal{RPT}$ symmetry,
as realized in Eq.~\eqref{eq:rho-Jz+RPT}.
The computer code has been developed accordingly,
and is planned to be published in the near future.

\begin{acknowledgments}
The authors are grateful to T.~Miyagi, N.~Hinohara and K.~Abe
for useful discussions.
This work is supported by the JSPS KAKENHI,
Grant Nos. JP24K07012 and JP25H00402.
Part of the numerical calculations has been performed
on Yukawa-21 at Yukawa Institute for Theoretical Physics in Kyoto University,
on SQUID at D3 Center in The University of Osaka,
on HITAC SR24000 at Digital Transformation Enhancement Council
in Chiba University.
\end{acknowledgments}


\end{document}